\documentclass[12pt,journal,letterpaper,twosides,onecolumn]{IEEEtran}

\usepackage[]{chapterbib} 
\usepackage{amsmath,amssymb,amscd,latexsym,dsfont,wasysym,bbm,stmaryrd}
\usepackage{float}
\usepackage{color}
\usepackage{graphicx,subfigure,balance}
\usepackage{multirow,multicol}
\usepackage{verbatim}
\usepackage{psfrag}
\usepackage{comment}
\usepackage{makeidx}
\usepackage[]{algorithm}
\usepackage{algorithmic}
\usepackage{cite}
\usepackage{slashbox}
\usepackage{arydshln} 
\newcommand{\tr}[1]{\textrm{#1}}

\newcommand{\tnr}[1]{{\textnormal{#1}}}
\newcommand{\mb}[1]{{\mathop{\boldsymbol{#1}}}}

\newcommand{\ov}[1]{\overline{#1}}

\newcommand{\secref}[1]{Sec.~\ref{#1}}

\newcommand{\theref}[1]{Theorem~\ref{#1}}

\newcommand{\ie}{i.e.,~} 		
\newcommand{\cf}{cf.~}		

\newcommand{\SET}[1]{\left\{#1\right\}}

\newcommand{\cd}{\cdot}
\newcommand{\ld}{\ldots}

\newcommand{\PR}[1]{\Pr\SET{#1}}       	
\newcommand{\pdf}{p}            			

\newcommand{\cdf}{F}            			

\newcommand{\siz}{1.4}  

\usepackage[acronym,nonumberlist]{glossaries} 
\usepackage{hyperref}
\newacronym[\glsshortpluralkey=PDFs,\glslongpluralkey=probability density functions]{pdf}{PDF}{probability density function}
\newacronym[\glsshortpluralkey=CDFs,\glslongpluralkey=cumulative density functions]{cdf}{CDF}{cumulative density function}
\newacronym[\glsshortpluralkey=CGFs,\glslongpluralkey=cumulant generating functions]{cgf}{CGF}{cumulant generating function}
\newacronym[\glsshortpluralkey=MGFs,\glslongpluralkey=moment generating functions]{mgf}{MGF}{moment generating function}
\newacronym[\glsshortpluralkey=PMFs,\glslongpluralkey=probability mass functions]{pmf}{PMF}{probability mass function}
\newacronym[]{cm}{CM}{coded modulation}
\newacronym[]{pam}{PAM}{pulse amplitude modulation}
\newacronym[]{bpsk}{BPSK}{binary phase shift keying}
\newacronym[]{qam}{QAM}{quadrature amplitude modulation}
\newacronym[]{psk}{PSK}{phase shift keying}
\newacronym[\glsshortpluralkey=LLRs,\glslongpluralkey=logarithmic likelihood ratios]{llr}{LLR}{logarithmic likelihood ratio}
\newacronym[]{map}{MAP}{maximum a posteriori}
\newacronym[]{ml}{ML}{maximum likelihood}
\newacronym[]{md}{MD}{multiuser diversity}
\newacronym[\glsshortpluralkey=MIs,\glslongpluralkey=mutual informations]{mi}{MI}{mutual information}
\newacronym[\glsshortpluralkey=GMIs,\glslongpluralkey=generalized mutual informations]{gmi}{GMI}{generalized mutual information}
\newacronym[]{bicm-gmi}{BICM-GMI}{BICM generalized mutual information}
\newacronym[]{awgn}{AWGN}{additive white Gaussian noise}
\newacronym[]{amc}{AMC}{adaptive modulation and coding}
\newacronym[]{sp}{SP}{set-partitioning}
\newacronym[]{gsm}{GSM}{global system for mobile communications}
\newacronym[]{edge}{EDGE}{enhanced data rates for GSM evolution}
\newacronym[]{3gpp}{3GPP}{3rd generation partnership project}
\newacronym[]{dvb}{DVB}{digital video broadcasting}
\newacronym[\glsshortpluralkey=PCCCs,\glslongpluralkey=parallel concatenated convolutional codes]{pccc}{PCCC}{parallel concatenated convolutional code}
\newacronym[\glsshortpluralkey=TCs,\glslongpluralkey=turbo codes]{tc}{TC}{turbo code}
\newacronym{ldpc}{LDPC}{low-density parity-check}
\newacronym[]{ofdm}{OFDM}{orthogonal frequency-division multiplexing}
\newacronym[]{bep}{BEP}{bit-error probability}
\newacronym[]{blep}{BLEP}{block-error probability}
\newacronym[]{sep}{SEP}{symbol-error probability}
\newacronym[]{ttcm}{TTCM}{turbo-trellis coded modulation}
\newacronym[]{uep}{UEP}{unequal error protection}
\newacronym[\glsshortpluralkey=CENCs,\glslongpluralkey=convolutional encoders]{cenc}{CENC}{convolutional encoder}
\newacronym[]{mimo}{MIMO}{multiple-input multiple-output}
\newacronym[\glsshortpluralkey=SNRs,\glslongpluralkey=signal-to-noise ratios]{snr}{SNR}{signal-to-noise ratio}
\newacronym[]{msb}{MSB}{most significant bit}
\newacronym[]{bcjr}{BCJR}{Bahl--Cocke--Jelinek--Raviv}
\newacronym[\glsshortpluralkey=SEDs,\glslongpluralkey=squared Euclidean distances]{sed}{SED}{squared Euclidean distance}
\newacronym[\glsshortpluralkey=EDs,\glslongpluralkey=Euclidean distances]{ed}{ED}{Euclidean distance}
\newacronym[\glsshortpluralkey=MEDs,\glslongpluralkey=minimum Euclidean distances]{med}{MED}{minimum Euclidean distance}
\newacronym[]{core}{CoRe}{constellation rearrangement}
\newacronym[]{msd}{MSD}{multistage decoding}
\newacronym[]{pdl}{PDL}{parallel decoding of the individual levels}
\newacronym[\glsshortpluralkey=GCs,\glslongpluralkey=Gray codes]{gc}{GC}{Gray code}
\newacronym[]{brgc}{BRGC}{binary-reflected Gray code}
\newacronym[]{nbc}{NBC}{natural binary code}
\newacronym[]{fbc}{FBC}{folded-binary code}
\newacronym[]{bsgc}{BSGC}{binary semi-Gray code}
\newacronym[]{msp}{MSP}{modified set-partitioning}
\newacronym[]{ssp}{SSP}{semi set-partitioning}
\newacronym[]{fhd}{FHD}{free Hamming distance}
\newacronym[]{mfhd}{MFHD}{maximum free Hamming distance}
\newacronym[]{ods}{ODS}{optimal distance spectrum}
\newacronym[]{iud}{i.u.d.}{independent and uniformly distributed}
\newacronym[]{ud}{u.d.}{uniformly distributed}
\newacronym[]{iid}{i.i.d.}{independent and identically distributed}
\newacronym[]{bico}{BICO}{binary-input continuous-output}
\newacronym[]{gh}{GH}{Gauss--Hermite}
\newacronym[]{sl}{SL}{Shannon limit} 

\newacronym[]{lhs}{l.h.s.}{left-hand side}
\newacronym[]{rhs}{r.h.s.}{right-hand side} 
\newacronym[]{ra}{RA}{resource allocation}
\newacronym[]{rr}{RR}{round-robin}
\newacronym[]{er}{ER}{equal-rate}
\newacronym[]{pf}{PF}{proportional fairness}
\newacronym[]{ts}{TS}{time-sharing}
\newacronym[]{sc}{SC}{superposition coding}
\newacronym[]{pf-ts}{PF-TS}{proportionally fair TS}
\newacronym[]{pf-sc}{PF-SC}{proportionally fair SC}
\newacronym[]{er-ts}{ER-TS}{equal-rate TS}
\newacronym[]{er-sc}{ER-SC}{equal-rate SC}

\newacronym[\glsshortpluralkey=BSs,\glslongpluralkey=base-stations]{bs}{BS}{base-station}
\newacronym[\glsshortpluralkey=MSs,\glslongpluralkey=mobile-stations]{ms}{MS}{mobile-stations}
\newacronym[]{kkt}{KKT}{Karush--Kuhn--Tucker} 
\newacronym[]{mcs}{MCS}{modulation/coding scheme} 
\newacronym[]{fec}{FEC}{forward error correction}
\newacronym[]{arq}{ARQ}{automatic repeat request}
\newacronym[]{harq}{HARQ}{hybrid ARQ}
\newacronym[]{tarq}{TARQ}{truncated HARQ}
\newacronym[]{ccharq}{CC-HARQ}{Chase combining HARQ}
\newacronym[]{irharq}{IR-HARQ}{incremental redundancy HARQ}
\newacronym[]{ack}{ACK}{positive acknowledgment}
\newacronym[]{nack}{NACK}{negative acknowledgment}
\newacronym[]{dp}{DP}{dynamic programming}
\newacronym[]{gp}{GP}{geometric programming}
\newacronym[]{csi}{CSI}{channel state information}
\newacronym[]{per}{PER}{packet error rate}
\newacronym[]{op}{OP}{outage probability}
\newacronym[]{spa}{SPA}{saddle-point approximation}
\newacronym[]{mrc}{MRC}{maximum ratio combining}

\newacronym[]{ir}{\tr{\tiny {IR}}}{IR}
\newacronym[]{cc}{\tr{\tiny {CC}}}{CC}
\newacronym[]{al}{\tr{\tiny {(AL)}}}{al}
\newacronym[]{ad}{\tr{\tiny{(AD)}}}{ad}
\newacronym[]{*}{\tr{\tiny {(*)}}}{*}
\newacronym[]{csik}{\mathsf{CSI}}{CSIk}

\newtheorem{Theorem}{Theorem}

\begin{document}




\title{Outage Minimization via Power Adaptation and Allocation for Truncated Hybrid ARQ}

\author{Mohammed Jabi,~Leszek Szczecinski,~Mustapha Benjillali,~and~Fabrice Labeau
\\
\thanks{%
The work was supported by the government of Quebec, under grant \#PSR-SIIRI-435.}%
\thanks{%
M. Jabi and L. Szczecinski are with INRS-EMT, Montreal, Canada. [e-mail: \{jabi,~leszek\}@emt.inrs.ca].}%
\thanks{%
M. Benjillali is with the Communications Systems Department, INPT, Rabat, Morocco. [e-mail: benjillali@ieee.org].}%
\thanks{%
F. Labeau is with McGill University, Montreal, Canada. [e-mail: fabrice.labeau@mcgill.ca].}%
 \thanks{%
Part of this work was submitted to {IEEE} International Conference on Communications, Sydney, Australia, 2014.} %
}%


\maketitle
\thispagestyle{empty}

\begin{abstract}
In this work, we analyze \gls{harq} protocols over the independent block fading channel. We assume that the transmitter is unaware of the \gls{csi} but has a knowledge about the channel statistics. We consider two scenarios with respect to the feedback received by the transmitter: {\it i)} ``conventional", one-bit feedback about the decoding success/failure (ACK/NACK), and {\it ii)} the multi-bit feedback scheme when, on top of ACK/NACK, the receiver provides additional information about the state of the decoder to the transmitter. In both cases, the feedback is used to allocate (in the case of one-bit feedback) or adapt (in the case of multi-bit feedback) the power across the HARQ transmission attempts. The objective in both cases is the minimization of the outage probability under long-term average and peak power constraints. We cast the problems into the \gls{dp} framework and solve them for Nakagami-$m$ fading channels. A simplified solution for the high \gls{snr} regime is presented using a \gls{gp} approach. The obtained results quantify the advantage of the multi-bit feedback over the conventional approach, and show that the power optimization can provide significant gains over conventional power-constant HARQ transmissions even in the presence of peak-power constraints.
\end{abstract}


\section{Introduction}\label{Sec:Introduction}
\IEEEPARstart{T}{o guarantee} reliable data transmissions over unreliable channels, two fundamental techniques are commonly used: \gls{fec} and \gls{arq} \cite {Lin03_Book}. In \gls{fec} schemes, error correcting codes are used to combat transmission errors. In \gls{arq} schemes, error detecting codes are used and retransmission is requested every time a \gls{nack} is sent to the transmitter via the feedback channel. An \gls{harq} scheme combines \gls{arq} and \gls{fec}, and provides better performances compared to each scheme alone~\cite{Shu84}. In typical \gls{harq} protocols, a retransmission request is repeated until the codeword is received without errors---in which case a \gls{ack} is sent on the feedback channel---or a maximum number of transmissions is reached; this particular case is called truncated \gls{harq} \cite{Malkamaki00,Liu04}. \gls{harq} schemes can be classified into two categories: the \gls{ccharq} \cite{Chase85}, where all retransmitted packets are identical, and the \gls{irharq} \cite{Hagenauer88} where each retransmission carries a different piece of the ``mother code'' that generates the complete coded version of the message.

In this paper, we design transmission schemes and power assignment strategies which minimize the outage probability subject to both peak power and long-term average power constraints for \gls{irharq} and \gls{ccharq} protocols in block fading channels. We analyze both cases when one bit \gls{ack}/\gls{nack} or multi-bits feedback is available at the transmitter. The multi-bit feedback scenario covers the case when the transmitter may obtain the \gls{csi} from the receiver through the feedback channel, but---due, e.g., to long communication/processing delays---the CSI is fully outdated (i.e., independent of the \gls{csi} in the subsequent transmissions).

To improve \gls{harq}'s performance, many power policies have been proposed in the litterature. In \cite{Tuninetti11b}, a power adaptation was proposed to increase the throughput in the case of a discretized \gls{csi}. An asymptotically optimal power control algorithm that attains the diversity limit in long-term static channels has been presented in \cite{Gamal06}. The case of long-term static channels was also studied in \cite{Su11}, where the authors determined the optimal power assignment strategy to minimize the total average transmission power subject to outage probability constraints. The optimization of power efficiency with a \gls{per} constraint was solved as a \gls{gp} problem in \cite {Hongbo09} for the case of space-time coded \gls{harq}, and in \cite {Chaitanya13} for \gls{ccharq} over independent Rayleigh block fading channels. In \cite{Chaitanya11}, the authors derived an optimal power allocation scheme which minimizes the packet drop probability under a total average transmit power constraint for \gls{irharq} with two transmissions. A suboptimal feedback and power adaptation rule was proposed for \gls{mimo} \gls{irharq} block fading channels in \cite{Nguyen12}, achieving the optimal outage diversity. 

The objective of this paper is to assess the value of the multi-bit feedback for power assignment schemes in \gls{harq}, and the main contributions of this work are the following:
\begin{enumerate}
\item We show how to use the well-known \gls{dp} methods \cite{Bertsekas05_book} to find the optimal power adaptation policies for truncated \gls{harq} in order to minimize the outage probability under constraints on peak and long-term average power. The method can be applied for both \gls{ccharq} and \gls{irharq} and for any channel with a continuous cumulative distribution function. Unlike \cite{Nguyen12}, where the proposed power strategies are sub-optimal in terms of outage performance, our power policies are optimal in terms of outage.
\item We show how to optimize the power-allocation policy for \gls{irharq} and \gls{ccharq} over Nakagami-$m$ fading channels. The optimal solutions are given in parametrized closed form for an arbitrary  number of transmissions. We note that only two transmissions were allowed in \cite{Chaitanya11}; in \cite {Chaitanya13} and \cite{Chaitanya11} only Rayleigh block fading channels were considered.
\item We present a simplified allocation policy for the high SNR regime obtained using the geometric programming (\gls{gp}) framework.
\item We provide numerical results for practically interesting wireless channel models, comparing the outage probability with various power adaptation/allocation methods.
\end{enumerate}
The rest of the paper is organized as follows. In~\secref{Sec:Model}, we introduce the adopted system model and, in~\secref{Sec:Optimization}, we define the optimization problem. We show the optimization method for power adaptation policies in~\secref{Sec:Pow_Adapt} and the power allocation is treated in~\secref{Sec:Pow_Alloc}. The optimal allocation for the high SNR regime is discussed in~\secref{Sec:feedback gain}. We provide numerical examples that illustrate the advantages obtained using the optimal power policies in~\secref{Sec:Num.Results}. Conclusions are drawn in~\secref{Sec:Conclusions}.

\section{System Model}\label{Sec:Model}

We consider a block-fading model where the channel between the transmitter and the receiver is varying (fading) randomly from one transmission to another but stays invariant during each of the transmissions, thus the signal received on the $k^\tr{th}$ \gls{arq} transmission round is given by
\begin{align}
 \mathbf{y}_{k}=\sqrt{\gamma_{k}\cd P_{k}(\gls{csik}_{k-1})} \cd  \mathbf{x}_{k}+  \mathbf{z}_{k},~~~\!\!k=1,...,K
\end{align}
where $ \mathbf{z}_{k}$ is a zero-mean, unit-variance Gaussian noise, $ \mathbf{x}_k$ is the unit-variance transmitted signal, 
$P_{k}(\gls{csik}_{k-1})\geq 0$ is the transmit power and is a function of the previous realization of the channel \mbox{$\gls{csik}_{k-1}=[\gamma_{1},\gamma_{2},...,\gamma_{k-1}]$}, where $\sqrt{\gamma_{k}}$ is  the instantaneous channel gain. Then, $\gamma_{k}$ has the meaning of an instantaneous nominal \gls{snr} (\ie which considers unitary power transmission) which is assumed to be perfectly known at the receiver but unknown to the transmitter. Thus, the transmitter cannot adjust the communication rate in the $k^{\tr{th}}$ transmission based on $\gamma_{k}$.

To recover from decoding errors, the  coded versions of a data packet are transmitted at most $K$ times. On top of the conventional one-bit signaling between the transmitter and the receiver (\gls{ack}/\gls{nack} messages), we also allow the receiver to send the \gls{csi} collected during unsuccessful transmission attempts back to the transmitter (entirely defined through SNR realizations $\gamma_{k}$) via the feedback channel (which is assumed error-free). The transmitter should be able to \emph{adapt} the transmit power during the $k^{\tr{th}}$ transmission attempt using the knowledge of $\gamma_{1},\ld,\gamma_{k-1}$. Thus, we will talk about {\it power adaptation} when the CSI are used to adjust the power used in each transmission. On the other hand, the {\it power allocation} covers the case when only the ``conventional'' one-bit feedback (ACK/NACK) is available. In this case, the transmitter responds to the reception of a NACK message by retransmitting the packet with a power that depends only on the transmission idex $k=1,\cdots,K$.

We assume that $\gamma_{k}$ can be modelled as \gls{iid} random variables with $\ov{\gamma}_{k}=\mathbb{E}_{\gamma_{k}}[\gamma_{k}]$, where  $\mathbb{E}_{\gamma}[\cd]$ denotes the mathematical expectation calculated with respect to $\gamma$. The independence of $\gamma_{k}$ can be justified by the practical scenario where the successive transmissions are not sent in adjacent time instants and, being sufficiently well separated, the realizations of the channel become---to all practical extent---independent \cite{Caire01}.

Most of the derivations will be done in abstraction of the particular fading distribution, but in numerical examples, we consider the popular Nakagami-$m$ fading profile. Hence, the channel normalized \gls{snr} $\gamma_{k}$ follows a gamma distribution with a \gls{pdf} $\pdf_{\gamma}(x)$ given by
\begin{align}
\pdf_{\gamma}(x)&=\frac{m^{m}}{\Gamma(m)\ov{\gamma}^{m}}x^{m-1}\tr{e}^{-m x/\ov{\gamma}}, \qquad x>0\nonumber,
\end{align}
and the \gls{cdf} $\cdf_{\gamma}(x)$ is given by
\begin{align}
\cdf_{\gamma}(x)=1-\frac{\Gamma\!\left(m,mx/\ov{\gamma}\right)}{\Gamma(m)},
\end{align}
where $\Gamma(x)$ and $\Gamma(s,x)$ denote respectively the gamma function and the upper incomplete gamma function.

We assume that the decoding is successful if the average accumulated mutual information at the receiver is larger than the overall transmission rate for \gls{irharq}. In the case of \gls{ccharq},  the decoding is successful if the accumulated SNR is larger than an SNR threshold. Thus, the decoding fails after $k$ transmissions with the probability

\begin{align}
f_{k}
&=\begin{cases} \label{eq.outage}
\PR {{\sum_{l=1}^{k}}\log \left(1{+}\gamma_{l}\cd P_{l}(\gls{csik}_{l-1}) \right)<R},&\text {for \gls{irharq}}\\
\PR  { \log \left(1{+} {\sum_{l=1}^{k}} \gamma_{l}\cd P_{l}( \gls{csik}_{l-1}) \right)<R},&\text{for  \gls{ccharq} }\\
\end{cases}\nonumber \\
&=\PR{I_{k}<i_{\tr{th}}} \\
&=\!\cdf_{I_{k}}(i_{\tr{th}})\!=\!\int_{0}^{i_{\tr{th}}}\pdf_{I_{k}}(x)\tr{d}x, \label{eq:cdf} 
\end{align}

where  
\begin{align}
\begin{cases}\label{eq.outage.App}
I_{k}={\sum_{l=1}^{k}} C_{l}, \quad i_{\tr{th}}=R & \text {for \gls{irharq}}\\
I_{k}={\sum_{l=1}^{k}} \sigma_{l}, \quad i_{\tr{th}}= \gamma_{\tr{th}}\!=\!2^{R}{-}1  & \text{for  \gls{ccharq}}\\
\end{cases},\\\nonumber
\end{align}
and $\sigma_{l}=\gamma_{l} \cd P_{l} (\gls{csik}_{l-1})$, \mbox{$C_{l}=\log \left(1+\gamma_{l}\cd P_{l}(\gls{csik}_{l-1}) \right)$} and $\pdf_{I_{k}}(x)$ is the \gls{pdf} of $I_{k}$.

With this notation, the scenarios we consider are defined as follows:
\begin{itemize}
\item Constant power (CO) \gls{harq}, where $P_{k}(\gls{csik}_{k-1})\equiv \bar{P}$, \ie the power is the same throughout retransmissions.

\item Power Allocation (AL), where the CSI feedback is ignored (or, simply Ðnot available) and the power varies solely as a function of the transmission's index, \ie $P_{k}(\gls{csik}_{k-1})\equiv \hat{P}_{k}\cd \mathbb{I}(I_{k-1}\leq i_{\tr{th}})$ where $\mathbb{I}(x)=1$ if $x$  is true, and $0$ otherwise, since the $k^{\tr{th}}$ transmission is necessary only if the previous $(k\!-\!1)$ transmissions were unsuccessful. Finding the scalars $\hat{P}_{k}$ is a problem of power allocation.
\item Power Adaptation (AD), where the power is modified in each transmission attempt using the CSI provided over the feedback channel. From \eqref{eq.outage}, the decoding error event in the $k^{\tr{th}}$ transmission depends uniquely on $I_{k-1}$ and $\gamma_{k}$ (which is unknown, and cannot be predicted from the previous CSI  $\gamma_{1}$,\ldots, $\gamma_{k-1}$ due to the independence assumption). Consequently, $I_{k-1}$ (which is a scalar representation of the vector $\gls{csik}_{k-1}$) is the only parameter eventually required to adapt the power $P_{k} (\gls{csik}_{k-1})$ via a scalar function
\begin{align}\label{eq:power_add}
P_{k}(\gls{csik}_{k-1})\equiv \tilde{P}_{k}(I_{k-1})\cd  \mathbb{I}(I_{k-1}\leq i_{\tr{th}}) ,\quad k=1,...,K
\end{align}
where $I_{0}\triangleq0$. Finding the function $\tilde{P}_{k}(I_{k-1})$ is a problem of power adaptation.
\end{itemize}

For simplicity, we assume that the transmitter has a perfect knowledge of $I_{k-1}$, that is, we ignore all eventual transmission and discretization errors. This assumption lets us know the maximum gain that can be achieved using information about the decoder's state contained in $I_{k-1}$.

\section{Optimization Problem}
\label{Sec:Optimization}

According to the reward-renewal theorem \cite{Zorzi96}, the long-term average consumed power is the ratio between the average transmit power between two consecutive renewals (sending a new data packet) $\mathbb{E}_{\gls{csik}_{K}}\Big[\mathcal{P}({\gls{csik}_{K})}\Big]$ and the expected number of transmissions $\mathbb{E}_{\gls{csik}_{K}}\Big[\mathcal{T}({\gls{csik}_{K}})\Big]$ needed to deliver the packet with up to $K$ transmission attempts \cite{Tuninetti11b}, \cite{Caire01}:
\begin{align}\label{P_avg}
\overline{P} \triangleq \frac{\mathbb{E}_{\gls{csik}_{K}}\Big[\mathcal{P}({\gls{csik}_{K}})\Big]}{\mathbb{E}_{\gls{csik}_{K}}\Big[\mathcal{T}({\gls{csik}_{K}})\Big]}= \frac{\displaystyle{\sum_{k=1}^{K} \mathbb{E}_{\gls{csik}_{k-1}}\!\!\Big[P_{k}({\gls{csik}_{k-1}})\Big]} }{\displaystyle{\sum_{k=0}^{K-1}} f_{k}},
\end{align}
where $f_{k}$ is the probability of a decoding failure after $k$ transmission attempts  given by \eqref{eq.outage}, and $ \mathbb{E}_{\gls{csik}_{k-1}}\!\!\Big[P_{k}({\gls{csik}_{k-1}})\Big]$  is the expected transmit power during the $k^{\tr{th}}$ transmission attempt, obtained by considering all the events yielding the $k^{\tr{th}}$ transmission, i.e., the event $I_{k-1} < i_{\tr{th}}$.

In this work, we aim at minimizing the outage probability $f_{K} $ with respect to the power policy $\{P_{k}({\gls{csik}_{k-1}})\}_{k=1}^{K}$ for a given long-term average power $\overline{P}_{\tr{max}}$, peak allowed power $P_{\tr{max}}$ and a transmission rate $R$. Taking \eqref{P_avg} into consideration, the optimization problem can be formulated as follows:


\begin{align} \label{eq:optimization}
 \underset{P_{1},P_{2}({\gls{csik}_{1}}),...,P_{K}({\gls{csik}_{K-1}})}{\min} f_{K},\qquad  \tnr{s.t.}\quad \begin{cases}
\overline{P} \leq \overline{P}_{\tr{max}}\\
0\leq P_{k}({\gls{csik}_{k-1}}) \leq P_{\tr{max}},~   1 \leq k \leq K
\end{cases}.
\end{align}

The problem  \eqref{eq:optimization} requires an  optimization over the scalar $P_1$ and the functions $P_{k}({\gls{csik}_{k-1}})$, so to solve it we will discretize the functions using $N$ equidistant points. Then we define the Lagrangian function {$L : \mathbf{R}_{+}^{N(K-1)+1}\!\times\!\mathbf{R} \rightarrow \mathbf{R}$} associated with the problem \eqref{eq:optimization} as
\begin{align}\label{eq:lagrange}
&L\left(P_{1},P_{2}({\gls{csik}_{1}}),...,P_{K}({\gls{csik}_{K-1}}),\lambda \right)=f_{K}+\lambda \cd \left(\sum_{k=1}^{K} \mathbb{E}_{\gls{csik}_{k-1}}\!\!\Big[P_{k}({\gls{csik}_{k-1}})\Big]-\overline{P}_{\tr{max}}\sum_{k=0}^{K-1}f_{k}\right)\!,
\end{align}
where we left implicit all power constraints \mbox{$0\leq P_{k}({\gls{csik}_{k-1}}) \leq P_{\tr{max}}$}. Without any loss of generality, we consider $\overline{P}_{\tr{max}}\!=\!1$ in what follows.

\section{outage-optimal power adaptation}
\label{Sec:Pow_Adapt}

For power adaptation \eqref{eq:power_add}, the expected transmit power during the $k^{\tr{th}}$ transmission is given by:
 \begin{align} \label{E.P.k}
\mathbb{E}_{\gls{csik}_{k-1}}\!\!\Big[P_{k}({\gls{csik}_{k-1}})\Big]&=\mathbb{E}_{I_{k-1}}\!\!\left[\tilde{P}_{k}(I_{k-1})\right] \nonumber \\
&=\int_{0}^{i_{\tr{th}}}\!\!\tilde{P}_{k}(x)\pdf_{I_{k-1}}\!(x)\tr{d}x.
\end{align}
Thus the Lagrangian function $L$ defined in \eqref{eq:lagrange} can be written as:
\begin{align} 
L\left(\tilde{P}_{1},\tilde{P}_{2}(I_{1}),\ld,\tilde{P}_{K}(I_{K-1}),\lambda \right)=f_{K}+\lambda \cd \left(\sum_{k=1}^{K} \mathbb{E}_{I_{k-1}}\!\!\left[\tilde{P}_{k}(I_{k-1})\right]- \sum_{k=0}^{K-1}f_{k}\right). \label{eq:lagrangP_ad}
\end{align}

To solve the primal problem in \eqref{eq:optimization}, it is difficult to use the \gls{kkt} conditions on the Lagrangian function \eqref{eq:lagrangP_ad} since it requires to solve analytically a system of an infinite number of equations where, in addition, closed form expressions of $f_k$ (for $2 \leq k \leq K$) are unknown. To overcome these difficulties, we can solve the dual problem. 

In the general case, the dual problem provides a solution which is a lower bound to the solution of \eqref{eq:optimization}; the difference between the lower bound and the true optimum is called the ``duality gap''. However, according to a result in \cite[Theorem 1]{Ribeiro10}, optimization problems with expectations over possibly non-convex functions of random variables in both objective and constraint functions have a zero duality gap, given that the \gls{pdf} of the random variable of interest has no points of strictly positive probability (i.e., its \gls{cdf} is continuous). For this reason, we express the outage probability and the average transmit power \eqref{E.P.k} as a function of the channel normalized SNR $\gamma_{k}$, $k=1,\ld,K$, with a continuous \gls{cdf}, indeed verifying the above requirement:
\begin{align}
f_{K} &=\PR{I_{K}<i_{\tr{th}} }=\mathbb{E}_{\gamma_{1},\gamma_{2},...\gamma_{K} }\!\!\left[\mathbb{I} (I_{K}<i_{\tr{th}} )\right]\!,
\end{align}
 and
\begin{align}\label{exp_p}
\mathbb{E}_{\gls{csik}_{K}}[P]&=\tilde{P}_{1} +\displaystyle{\sum_{k=2}^{K} \mathbb{E}_{\gamma_{1},\gamma_{2},...\gamma_{k-1} }\!\!\left[\tilde{P}_{k}(I_{k-1})\right]}\nonumber\\
&=\tilde{P}_{1} + \mathbb{E}_{\gamma_{1},\gamma_{2},...\gamma_{K-1} }\!\!\left[\displaystyle{\sum_{k=2}^{K} \tilde{P}_{k}(I_{k-1})}\right]\!.
\end{align}

According to  \cite[Theorem 1]{Ribeiro10}, the objective and constraint functions must be expectations over (possibly non-convex) functions of random variables. However, $\tilde{P}_{1}$ is independent of any random variable. So we introduce a sub-optimization problem for each value of $\tilde{P}_{1}>0$
\begin{align} \label{eq:sub-optimization}
~\hat{f}_{K}(\tilde{P}_{1}) \triangleq \underset{\tilde{P}_{2}(I_{1}),...,\tilde{P}_{K}(I_{K-1})}{\min} f_{K},
~~~\tnr{s.t.} \begin{cases}
\displaystyle{\sum_{k=2}^{K} \mathbb{E}_{I_{k-1}}\!\!\left[\tilde{P}_{k}(I_{k-1})\right]-\sum_{k=2}^{K-1}f_{k}} \leq 1+f_{1}-\tilde{P}_{1}\\
0\leq \tilde{P}_{k}\left(I_{k-1}\right)\leq P_{\tr{max}}, \quad \text{for} \quad   1 \leq k \leq K
\end{cases}\!\!\!\!.
\end{align}
The optimal solution of \eqref{eq:optimization} is then given by
\begin{align}\label{eq:g-optimization}
\underset{\tilde{P}_{1}}{\min} ~\hat{f}_{K}(\tilde{P}_{1}).
\end{align}
Defining the Lagrange dual function $d:\mathbf{R}_{+}\times\mathbf{R} \rightarrow \mathbf{R}$ as
\begin{align} \label{eq:L.dual}
d( \tilde{P}_{1},\lambda){\triangleq}\underset{\tilde{P}_{2}(I_{1}),\ld,\tilde{P}_{K}(I_{K-1})}{\min}\!\!L\!\left(\!\tilde{P}_{1},\!\tilde{P}_{2}(I_{1}),\ld,\!\tilde{P}_{K}(I_{K-1}\!),\lambda\!\right)\!\!,
\end{align}\label{eq:optimization.dual}
the dual optimization problem is thus given by
\begin{align}\label{eq:optimization.dual} 
D(\tilde{P}_{1})=\underset{\lambda\geq 0}{\min} ~ d( \tilde{P}_{1},\lambda).
\end{align}

Note that, since the problem \eqref{eq:sub-optimization} and its dual \eqref{eq:optimization.dual} have a zero duality gap, we can guarantee that $D(\tilde{P}_{1})\equiv\hat{f}_{K}(\tilde{P}_{1})$ for $\tilde{P}_{1}>0$. 

Finally, \eqref{eq:L.dual} can be rewritten in a recursive form characteristic of dynamic programming optimization (DP): 
\begin{align}
d( \tilde{P}_{1},\lambda)&=J_{1}(I_{0} ) \nonumber \\
J_{1}(I_{0}) &=\Big\{-\lambda \cd \mathbb{E}_{\gamma_{1}}\!\!\left[\mathbb{I} (I_{1}<i_{\tr{th}})\right]+\lambda \cd \tilde{P}_{1} +\mathbb{E}_{\gamma_{1}}\!\!\left[J_{2} ( I_{1})\right] \Big \}  \label{D.P.first} \\  
J_{2}(I_{1}) &=\underset{\tilde{P}_{2}(I_{1})}{\min} \Big\{-\lambda \cd \mathbb{E}_{\gamma_{2}}\!\!\left[\mathbb{I} (I_{2}<i_{\tr{th}})\right]+\lambda \cd \tilde{P}_{2}(I_{1})+\mathbb{E}_{\gamma_{2}}\!\!\left[J_{3} (I_{2})\right] \Big\}\\
& \vdots   \nonumber \\    
J_{k}(I_{k-1})&=\underset{\tilde{P}_{k}(I_{k-1})}{\min}\Big\{-\lambda\cd\mathbb{E}_{\gamma_{k}}\!\!\left[\mathbb{I} (I_{k}<i_{\tr{th}})\right]+\lambda\cd \tilde{P}_{k}(I_{k-1})+\mathbb{E}_{\gamma_{k}}\!\!\left[J_{k+1} (I_{k})\right] \Big\}\\
& \vdots  \nonumber   \\  
 J_{K}(I_{K-1})&=\underset{\tilde{P}_{K}(I_{K-1})}{\min} \Big\{ \lambda \cd \tilde{P}_{K}(I_{K-1})+\mathbb{E}_{\gamma_{K}}\!\!\left[\mathbb{I} (I_{K}<i_{\tr{th}})\right] \Big\}, \label{D.P.last}
\end{align}
where $I_{k}$ is a function of $I_{k-1}$, $\tilde{P}_{k}(I_{k-1})$, and $\gamma_{k}$ in the form
\begin{align}
I_{k}=
\begin{cases}
I_{k-1}+\log\!\left(1+\gamma_{k} \tilde{P}_{k}(I_{k-1})\right), &~\text{for \gls{irharq}}\\
I_{k-1}+\gamma_{k} \tilde{P}_{k}(I_{k-1}),&~\text{for \gls{ccharq}}
\end{cases}.
\end{align}
For a given $I_{k}$---noting that $I_{k}\!\!\in\!\!\left[0,i_{\tr{th}}\right]$ should be discretized over $N$ points---we can optimize the value of the function $\tilde{P}_{k}(I_{k-1})$ provided that the function $J_{k+1}(I_{k})$ is known. Thus, the global optimization of the possibly non-convex problem in \eqref{eq:sub-optimization} over the set of $N^{K-1}$ values is reduced to a series of $(K-1)\cd N$ one-dimensional optimizations thanks to the \gls{dp} formulation equations in \eqref{D.P.first}-\eqref{D.P.last}. 



\subsection{ Radio silence}
In this context of \gls{irharq} transmissions, we have
\begin{align}
\mathbb{E}_{\gamma_{k}}\!\!\Big[\mathbb{I} (I_{k}<i_{\tr{th}})\Big]=\cdf_{\gamma_{k}}\left(\frac{2^{R-I_{k-1}}-1}{\tilde{P}_{k}(I_{k-1})} \right).
\end{align}
The condition to guarantee a minimum in the last \gls{dp} step is that the derivative of the function under minimization in \eqref{D.P.last} equals zero, i.e.,
\begin{align}
u(\tilde{P}_{K})&\triangleq\lambda-\frac{2^{R-I_{K-1}}-1}{\tilde{P}_{K}^2}\cdot p_{\gamma_{K}}\!\!\left(\frac{2^{R-I_{K-1}}-1}{\tilde{P}_{K}} \right)\nonumber\\
&=\lambda-\frac{1}{2^{R-I_{K-1}}-1}\cdot q\!\left(\frac{2^{R-I_{K-1}}-1}{\tilde{P}_{K}}\right)=0, \label{zero.der}
\end{align}
where it is easy to show that $q(x)\!\triangleq\!x^{2} p_{\gamma_{K}}\!(x)$ satisfies $q(x)>0$, $q(0)=0$, and $q(\infty)=0$, and hence, $q(x)$ has a maximum $q_\tr{max}=\max_{x}q(x)$. Since the derivative $u(0)=\lambda$ and $u(\tilde{P}_{K})$ must be locally non-increasing around $\tilde{P}_{K}=0$ (i.e., $u'(0)\leq 0$), the solution of \eqref{zero.der} does not exist if $\lambda\cdot(2^{R-I_{K-1}}-1)>q_\tr{max}$;  meaning that the minimum is obtained by setting $\tilde{P}_{K}=0$ and yielding $J_{K}(I_{K-1})=1$. When $\lambda\!\cdot\!(2^{R-I_{K-1}}\!-\!1)<q_\tr{max}$, $u(\tilde{P}_{K})$ has at least two zeros\footnote{In the case of a Rayleigh fading channel, there are exactly two zeros: one corresponds to the local maximum, and the other one to the local minimum.} and the optimal solution corresponds to the point where the second derivative is positive.

\begin{figure}[th]
\psfrag{xlabel}[c][c][\siz]{$I_{k-1}$}
\psfrag{ylabel}[c][c][\siz]{$\tilde{P}_{k}(I_{k-1})$}
\psfrag{first}[l][l][\siz]{$\tilde{P}_{1}$}
\psfrag{2power}[l][l][\siz]{$\tilde{P}_{2}(I_{1})$}
\psfrag{3power}[l][l][\siz]{$\tilde{P}_{3}(I_{2})$}
\psfrag{4power}[l][l][\siz]{$\tilde{P}_{4}(I_{3})$}
\begin{center}
\scalebox{0.5}{\includegraphics[width=1\linewidth]{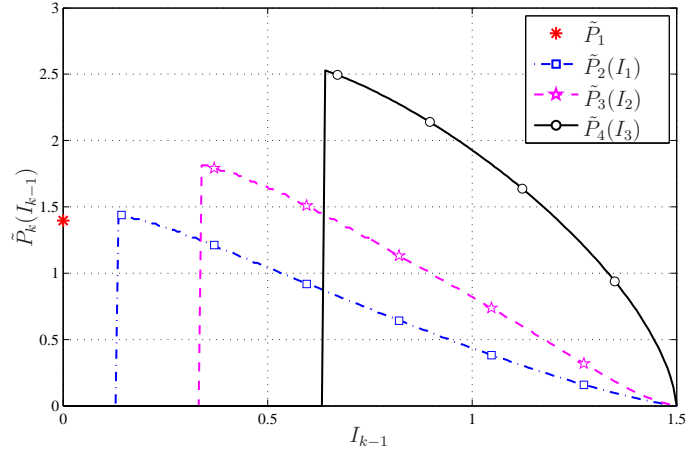}}
\caption{Optimized adaptation policies $\tilde{P}_{k}(I_{k-1})$ for the case of IR-HARQ in Nakagami-$m$ fading model, when $m=2$,  $ K=4$, $R=1.5$ and $\ov{\gamma}=-4$ dB. The ``radio silence'' means that if the accumulated mutual information after the $k^{\tr{th}}$ transmission is less than a specific threshold $i_{0,k}$ (\ie if $I_{k}<i_{0,k}$ where $i_{0,1}\approx 0.12$, $i_{0,2}\approx 0.33$, $i_{0,3}\approx 0.63$ in this example), the HARQ process decides to use zero power for all reminder transmissions, \ie $P_{l}(I_{l-1})=0,~k<l\leq K$.} 
\label{power.IR}
\end{center}
\end{figure}

In Fig.~\ref{power.IR}, we show the adaptation policy $\tilde{P}_{k}(x)$ in a case of \gls{irharq} with $K = 4$. As we see, the optimal solution requires a ``radio silence'', that is, knowing in the $k^{\tr{th}}$ transmission that the accumulated mutual information at the receiver is below a threshold $i_{0,k-1}$, the transmitter decides to stay silent (zero transmit power) until the maximum number of transmissions is attained. This ``silence time'' guarantees that the power is ``saved'' when the transmitter does not have a ``reasonable  hope'' of successfully terminating the transmission.
\subsection{ Outage calculation}

To calculate the outage probability, we use \eqref{eq:cdf} where the \gls{cdf} of $I_{k}$ is $\cdf_{I_{k}}(x)$ and taking into consideration that $\tilde{P}_{k}(x)=0$ for $x \in [0,i_{0,k-1}]$, we obtain
\begin{align}\label{F.I.k}
\cdf_{I_{k}}(x)=&\PR{I_{k-1}+\log(1+\gamma_{k} \cd \tilde{P}_{k}(I_{k-1}))<x}\nonumber\\
&=\begin{cases}
\cdf_{I_{k-1 }}\!(x),&\text {if~}  x<i_{0,k-1} \\
&\\
\cdf_{I_{k-1 }}\!(i_{0,k-1})+\displaystyle{\int_{i_{0,k-1} }^{x}\!\!\cdf_{\gamma}\!\!\left(\frac{2^{x-y}-1}{\tilde{P}_{k}(y)}\right)\cdot\pdf_{I_{k-1}}(y)~\!\tr{d}y},&\text {if~} x>i_{0,k-1} \\
\end{cases},
\end{align}
which depends on the \gls{pdf} $\pdf_{I_{k-1}}(y)$ of $I_{k-1}$.

The differentiation of \eqref{F.I.k} yields a recursive relationship for the \gls{pdf}
\begin{align}\label{p.I.k}
\pdf_{I_{k}}(x)=
\begin{cases}
\pdf_{I_{k-1 }}(x),&\text {if~} x<i_{0,k-1}\\
&\\
\int_{i_{0,k-1} }^{x} \frac{\log(2)  2^{x-y} }{\tilde{P}_{k}(y)}\pdf_{I_{k-1}}\!(y)\cdot\pdf_{\gamma}\!\!\left(\frac{2^{x-y}-1}{\tilde{P}_{k}(y)}\right)\tr{d}y&\text {if~} x>i_{0,k-1} 
\end{cases},
\end{align}
where
\begin{align}\label{p.I.1}
\pdf_{I_{1}}\!(x)=\frac{\log(2) 2^{x}}{\tilde{P}_{1}}\cdot \pdf_{\gamma}\!\left(\frac{2^{x}-1}{\tilde{P}_{1}}\right)\!.
\end{align}

Considering $\mathbb{E}_{\gamma_{k}}\!\!\left[\mathbb{I} (I_{k}<i_{\tr{th}}\right]=\cdf_{\gamma_{k}}\!\!\left(\!\frac{\gamma_{\tr{th}}-I_{k-1} }{\tilde{P}_{k}(I_{k-1})}\!\right)$, the same analysis as \eqref{F.I.k} and \eqref{p.I.k} can be done  in the case of \gls{ccharq}. Thus, 
\begin{align}\label{F.I.k.2}
\cdf_{I_{k}}(x)&=\PR{I_{k-1}+\gamma_{k} \cd \tilde{P}_{k}(I_{k-1})<x}\nonumber\\
&=
\begin{cases}
\cdf_{I_{k-1 }}(x), & \text {if~~} x<i_{0,k-1}\\
&\\
\cdf_{I_{k-1 }}(i_{0,k-1})+\displaystyle{\int_{i_{0,k-1} }^{x} \!\!\cdf_{\gamma}\!\!\left(\frac{x-y}{\tilde{P}_{k}(y)}\right)\cdot\pdf_{I_{k-1}}(y)~\!\tr{d}y}, & \text {if~~} x>i_{0,k-1}\\
\end{cases},
\end{align}
which again depends on the \gls{pdf} $\pdf_{I_{k-1}}(y)$ of $I_{k-1}$.

Differentiation of \eqref{F.I.k.2} yields the recursive relationship for the pdf
\begin{eqnarray}
  \pdf_{I_{k}}(x)&=&
  \begin{cases}
\pdf_{I_{k-1 }}(x) & \text {if} \qquad  x<i_{0,k-1} \\
&\\
 \int_{i_{0,k-1} }^{x} \frac{1 }{\tilde{P}_{k}(y)}\pdf_{I_{k-1}}(y)\pdf_{\gamma}\left(\frac{x-y}{\tilde{P}_{k}(y)}\right)   \tr{d}y, & \text {if} \qquad  x>i_{0,k-1} \\
\end{cases},
\end{eqnarray}
where
\begin{align}
\pdf_{I_{1}}(x)=\frac{1}{P_{1}}\pdf_{\gamma}(\frac{x}{P_{1}}).
\end{align}
\section{outage optimal power  allocation}
\label{Sec:Pow_Alloc}

In this section, we consider the problem of optimal power allocation \Big(\ie $P_{k}(\mathcal{{CSI}}_{k-1})\equiv \hat{P}_{k}\cd \mathbb{I}(I_{k-1}\leq i_{\tr{th}})$\Big). The expected power consumed in the $k^{\tr{th}}$ transmission attempt is given by
\begin{align} \label{E.P.K.Allo}
\mathbb{E}_{\gls{csik}_{k}}[P]\!=\!\mathbb{E}_{{I}_{k-1}}[ \hat{P}_{k}\cd \mathbb{I}(I_{k-1}\leq i_{\tr{th}})]\!=\!\hat{P}_{k}\cd f_{k-1},
\end{align}
and the long-term average power \eqref{P_avg} by
\begin{align}\label{P_avg_all}
\overline{P}=\frac{\displaystyle{\sum_{k=1}^{K} \hat{P}_{k}\cd f_{k-1}} }{\displaystyle{\sum_{k=0}^{K-1}} f_{k}}.
\end{align}

Thus, the Lagrangian function $L$ defined in (\ref{eq:lagrange}) can be expressed as
\begin{align}\label{eq:Allocation}
L(\hat{P}_{1},\hat{P}_{2},\ld,\hat{P}_{K},\lambda)= f_{K}+\lambda\cd\left(\sum_{k=1}^{K} \hat{P}_{k}  f_{k-1} - \sum_{k=0}^{K}f_{k} \right).
\end{align}

To cast (\ref{eq:Allocation}) into the \gls{dp} formulation, we have to find the ``states'' $S_{k}$ such that: (i) $f_{k}$ may be calculated from $S_{k}$, and (ii) state $S_{k+1}$ may be obtained from $S_{k}$ and $\hat{P}_{k+1}$. Because the closed form expressions
of $f_{k}$ are unknown, we use an accurate approximation to express $f_{k}$ in terms of  $\{\hat{P}_{l} \}_{l=1}^{k}$; as shown in the Appendix A, we can obtain the following relationship:
\begin{align}\label{A.S.R}
f_{k}\approx \frac{h_{k}}{\hat{P}_k^{m_{k}}}\cd f_{k-1}, \quad \text{for} \quad 1 \leq k \leq K,
\end{align}
where the parameter $h_{k}$ is independent from $\hat{P}_{k}$ and $m_{k}$ is the parameter of the Nakagami-$m$ channel at the $k^{\tr{th}}$ transmission. Considering $f_{0}=1$, the optimization problem in \eqref{eq:Allocation} can be reformulated recursively as follows
\begin{eqnarray}
L(\hat{P}_{1},\hat{P}_{2},...,\hat{P}_{K},\lambda)&=&J_{1}{(f_{0})} \nonumber \\
J_{1}{(f_{0})}&=& \lambda \cd (\hat{P}_{1}-1)\cd f_{0}+J_{2}{(f_{1})} \label{DP.Allo.1} \\
J_{2}{(f_{1})}&=& \lambda \cd (\hat{P}_{2}-1) \cd f_{1}+J_{3}{(f_{2})} \\
&&  \vdots   \nonumber  \\
J_{K-1}{(f_{K-1})}&=& \lambda \cd (\hat{P}_{K}-1) \cd f_{K-1}+ f_{K}. \label{DP.Allo.K}
\end{eqnarray}

From the \gls{kkt} necessary conditions, and starting with \eqref{DP.Allo.K}, we find a unique (therefore, the optimal) solution as
\begin{align}
\hat{P}_{k}
&=\begin{cases}
\min \left\{ \left(\frac{m_{k}\cd g_{k+1} \cd h_{k}}{\lambda} \right)^{\frac{1}{m_{k}+1}} , P_{\tr{max}} \right\},& \text {for}~1\leq k \leq K-1 \\
\min\left\{\left(\frac{m_{K}\cd h_{K}}{\lambda} \right)^{\frac{1}{m_{K}+1}} , P_{\tr{max}} \right\},& \text {for} ~ k=K
\end{cases} ,
\end{align}
where
\begin{align}
g_{k}&\triangleq\begin{cases}
\lambda \cd (\hat{P}_{k}-1)+\frac{g_{k+1} \cd h_{k}}{\hat{P}_{k}^{m_{k}}}, & \text {for}~1\leq k \leq K-1 \\
\lambda \cd (\hat{P}_{K}-1)+\frac{h_{K}}{\hat{P}_{K}^{m_{K}}}, & \text {for} ~ k=K
\end{cases}.
\end{align}


\section{Approximate solutions}
\label{Sec:feedback gain}
We target the high SNR regime, where typically \mbox{$\underset{{\ov{\gamma}\rightarrow\infty}}{\lim}f_k=0$} for $k \in \{1,...,K\}$. Thus, the long term average power (defined in \eqref{P_avg_all})  can be expressed in the high SNR regime as:
\begin{align}\label{p_avg_high_SNR}
\overline{P} \approx {\sum_{k=1}^{K} \hat{P}_{k}\cd f_{k-1}}.
\end{align}
We note that \cite {Chaitanya13} and  \cite{Chaitanya11} define the long term average power as in \eqref{p_avg_high_SNR} even if it is only a valid approximation in high SNR regime. Thus, the optimization problem \eqref {eq:optimization} can be rewritten in the case of power allocation as:\\
\begin{align}\label{eq:simpl.Opti.P.All}
 \underset{\hat{P}_{1},\hat{P}_{2},...,\hat{P}_{K}}{\min} f_{K},~~~~ \tnr{s.t.} ~~\displaystyle{\sum_{k=1}^{K} \hat{P}_{k}\cd f_{k-1}}  \leq 1,
\end{align}
where we assume that $ P_{\tr{max}}=\infty$ and $\overline{P}_{\tr{max}} =1$.

As shown in Appendix A, for Nakagami-$m$ fading channel, a unified approximation of the outage probability $f_{K}$ and the expected transmit power for both  \gls{irharq} and \gls{ccharq} can be written as 
\begin{align}\label{eq:A.f.K.Al}
f_{K} \approx A_{K} \cd \prod_{k=1}^{K} \hat{P}_{k}^{-m},
\end{align}
and
\begin{align}\label{eq:A.E.K.Al}
\hat{P}_{k}\cd f_{k-1} \approx A_{k-1}   \cd \hat{P}_{k}  \prod_{l=1}^{k-1} \hat{P}_{l}^{-m}.
\end{align}
where $A_{0}=1$, $A_{k}$ is defined in \eqref {eq_A_IR} and  \eqref{eq_A_CC} for the case of  \gls{irharq} and \gls{ccharq} respectively.

Thus, the optimization problems \eqref{eq:simpl.Opti.P.All} can be written in the standard primal form of geometric programming problems \cite{GPbook}, \cite{Luptacik}, \cite{Boyd04_Book}:
\begin{align} \label{eq:GP.op}
\underset{{\hat{P}_{1},\hat{P}_{2},...,\hat{P}_{K}}}{\min} \Big\{  A_{K} \cd \prod_{k=1}^{K} \hat{P}_{k}^{-m} \Big\},~~~\tnr{s.t.}~\sum_{k=1}^{K} A_{k-1} \cd \hat{P}_{k}  \prod_{l=1}^{k-1} \hat{P}_{l}^{-m}  \leq 1, 
\end{align}
As shown in Appendix B, the optimal solution of \eqref{eq:GP.op} is given by
\begin{align}\label{eq:GP.Sol.1}
f_{K}=\left(\lambda(\mb{\delta}^{*}) \right)^{\lambda(\mb{\delta}^{*})} \cd A_{K} \cd \prod_{k=2}^{K+1} \left( \frac{A_{k-2}}{\delta_{k}^{*}} \right)^{\delta_{k}^{*}},
\end{align}
where $\boldsymbol{\delta}^{*}=[\delta_1^{*}, \ld, \delta_{K+1}^{*}]$ and 
\begin{align} 
\delta_{k}^{*}=
\begin{cases} 
1, & \text{for} ~ k=1\\
m\cd (m+1)^{(K+1-k)}, & \text{for} ~ k\in\{2,...,K+1\}
\end{cases},
\end{align}
and thus $\lambda(\mb{\delta}^{*})=(m+1)^{K}-1$, \cf \eqref{eq:lambda}. Therefore, the optimal power policy corresponding to the optimization problem \eqref{eq:simpl.Opti.P.All} is given by \eqref{eq:x1.GP}
\begin{align}
P_{k}=
\begin{cases}
\displaystyle{\frac{{\delta}_{2}^{*}}{\lambda(\mb{\delta}^{*}) \cd A_{0}}},& \text{if}~ k=1\\
\displaystyle{\frac{{\delta}_{i+1}^{*}}{\lambda(\mb{\delta}^{*}) \cd A_{i-1}\cd \prod_{j=1}^{i-1} \left(P_{j}^{*}\right)^{-m}}}, & \text{if}~k\in\{2,..,K\}
\end{cases}.
\end{align}

On the other hand, the diversity is defined as \cite{Zheng03}
\begin{align}
\mathcal{D}=-\lim_{\ov{\gamma}\rightarrow \infty} \frac{\log (f_{K})}{\log(\ov{\gamma} )}.
\end{align}
In the $A_{k}$'s expressions,  the exponent of $\ov{\gamma}$ is equal to $-mk$. Thus, according to \eqref{eq:GP.Sol.1} the diversity $\mathcal{D}$ for power allocation is given by
\begin{eqnarray}
\mathcal{D}&=&Km{\delta}_{1}+\sum_{k=2}^{K+1} (k-2)m{\delta}_{k}\nonumber\\
&=&Km {\delta}_{1}+0m{\delta}_{2}+1 m{\delta}_{3}+...+(K-1) m{\delta}_{K+1} \nonumber\\
&=&Km+m^{2} [ (m+1)^{K-2}+2(m+1)^{K-3}+...+(K-2)(m+1)+(K-1) ] \nonumber\\
&=&Km+m^{2}\cd K\cd \left (1+(m+1)+...+(m+1)^{K-2}\right)-1-2\cd(m+2)-...-(K-1)\cd(m+1)^{K-2} \nonumber\\
&=& (m+1)^{K}-1
\end{eqnarray}

We note that the same diversity value was obtained in \cite{Nguyen12} where, in addition, it was proven that both multi-bit feedback (i.e., adaptation) and single-bit feedback (i.e., allocation) have the same diversity gain for an infinite constellation size; this is confirmed by our results. On the other hand, the diversity of constant power HARQ is given by $\mathcal{D}=Km$.

\section{numerical examples}
\label{Sec:Num.Results}

\begin{figure}[h!] 
\psfrag{xlabel}[c][c][\siz]{$\ov{\gamma}$\! {[dB]}}
\psfrag{ylabel}[c][c][\siz]{$f_{K}$}
\psfrag{K=2}[l][l][\siz]{$K=2$}
\psfrag{K=4}[l][l][\siz]{$K=4$}
\psfrag{Constant Alloc}[l][l][\siz]{CO}
\psfrag{DP Alloc unb}[l][l][\siz]{AL with $P_{\tr{max}}=\infty$ }
\psfrag{DP Alloc with Pma}[l][l][\siz]{AL with $P_{\tr{max}}=5$}
\psfrag{DP Adap unb}[l][l][\siz]{AD with $P_{\tr{max}}=\infty$ }
\psfrag{DP Adap with Pma}[l][l][\siz]{AD  with $P_{\tr{max}}=5$ }
\begin{center}
\scalebox{0.5}{\includegraphics[width=1\linewidth]{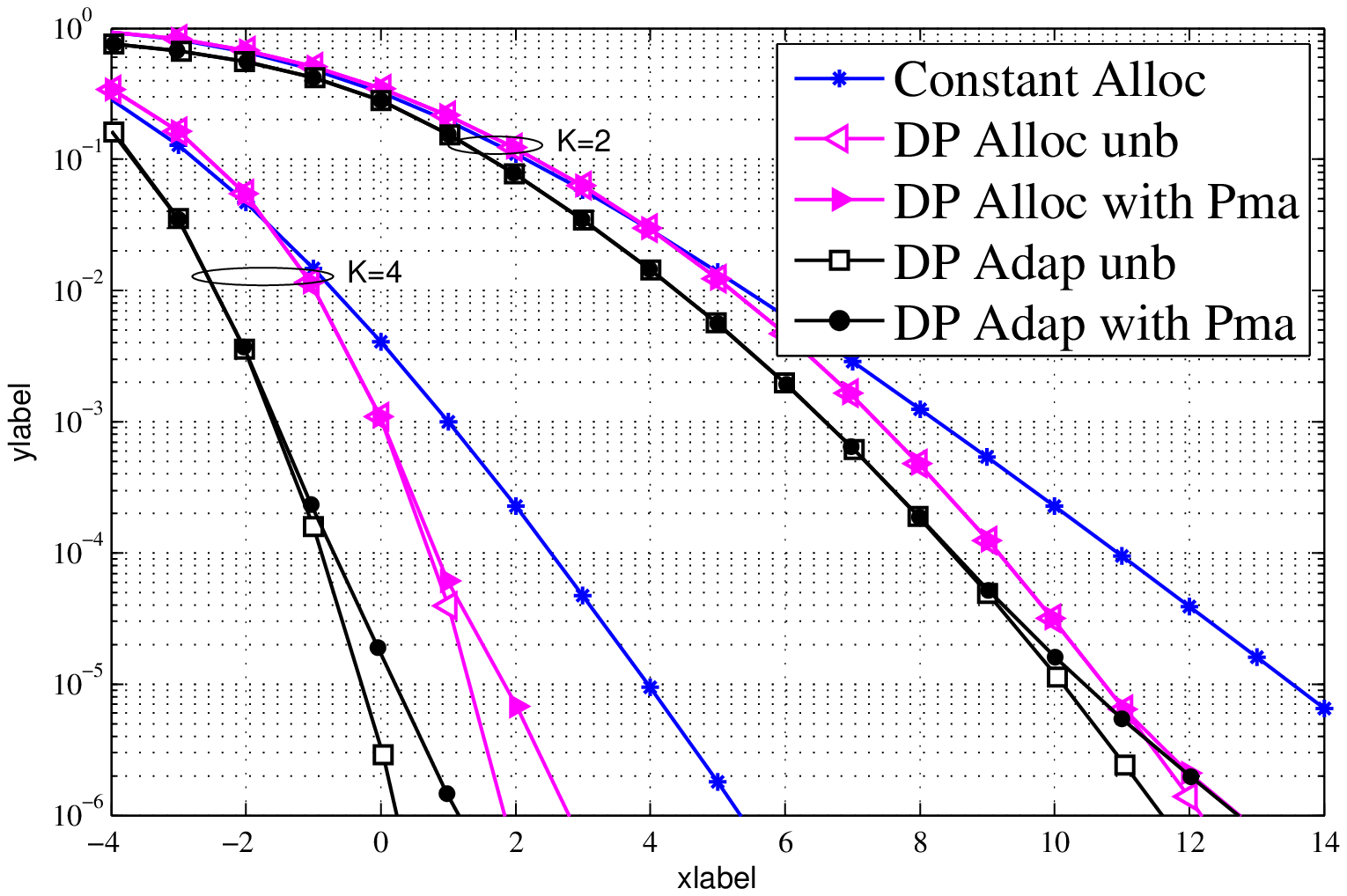}}
\\
(a)
\\
\scalebox{0.5}{\includegraphics[width=1\linewidth]{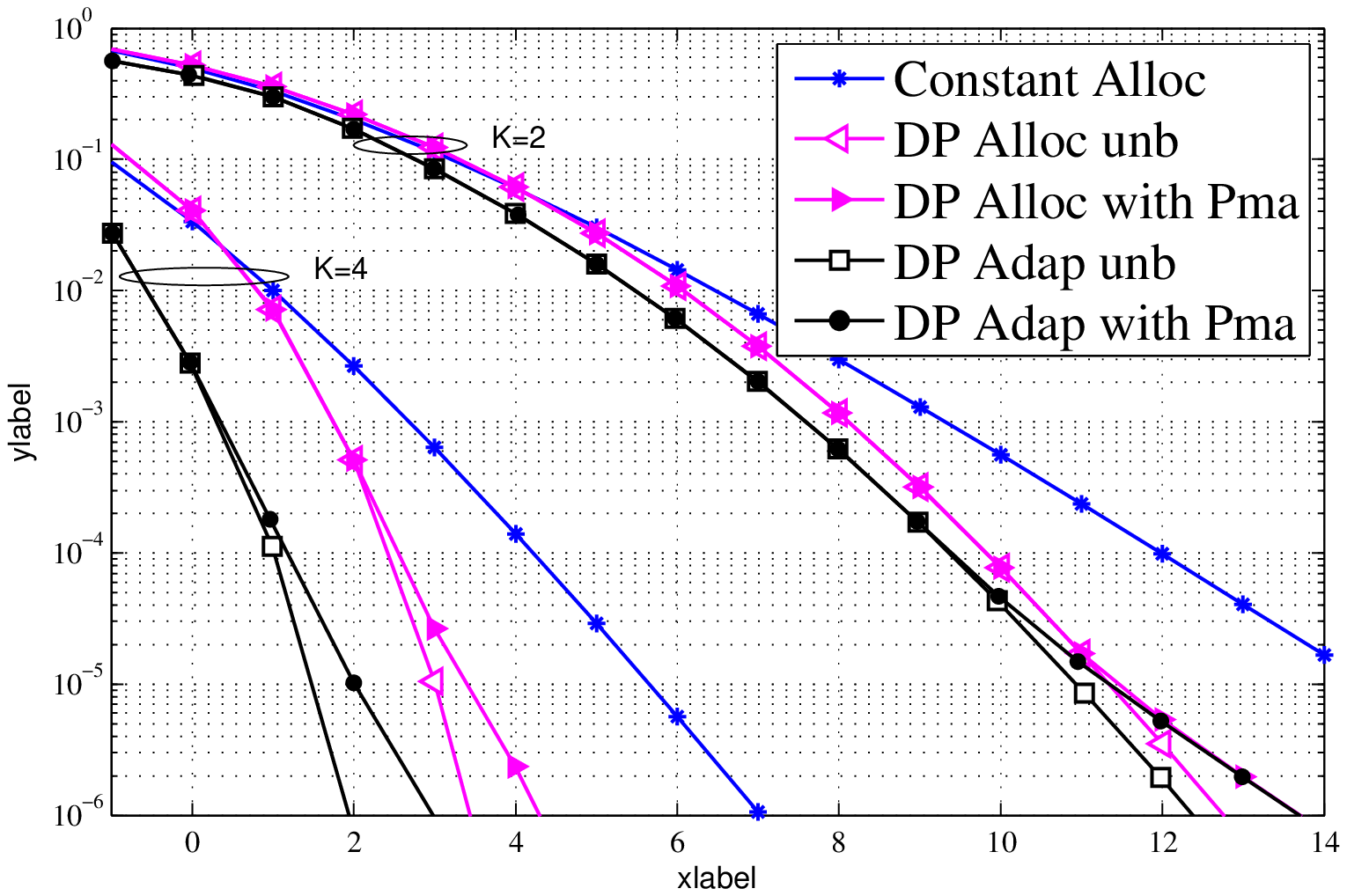}}
\\
(b)
\caption{Optimized outage probability when using the optimized power adaptation (AD) and allocation (AL) compared with the outage probability of constant-power transmission (CO) (i.e., $P_k=\ov{P}=1$, $\forall k$) in the case of (a) IR-HARQ, and (b) CC-HARQ. $K=2, 4$ and Nakagami-$m$ fading with $m=2$; $R=1.5$. Unconstrained peak (i.e., $P_{\tr{max}}=\infty$) and constrained peak  (i.e., $P_{\tr{max}}=5$) cases are shown.}\label{fig.outage.IR}
\end{center}
\end{figure}

\begin{figure}[h!]
\psfrag{xlabel}[c][c][\siz]{$\ov{\gamma}$\! {[dB]}}
\psfrag{ylabel}[c][c][\siz]{$\hat{P}_{k}$}
\psfrag{ubondddd}[l][l][\siz]{$P_{\tr{max}}=\infty$}
\psfrag{bond}[l][l][\siz]{$P_{\tr{max}}=5$}
\psfrag{first}[l][l][\siz]{$\hat{P}_{1}$}
\psfrag{2 po}[l][l][\siz]{$\hat{P}_{2}$}
\psfrag{3 po}[l][l][\siz]{$\hat{P}_{3}$}
\psfrag{4 po}[l][l][\siz]{$\hat{P}_{4}$}
\begin{center}
\scalebox{0.5}{\includegraphics[width=1\linewidth]{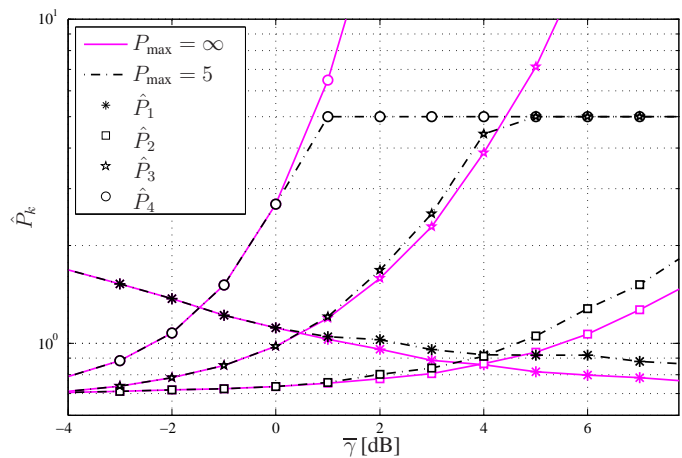}}
\\
(a)
\\
\scalebox{0.5}{\includegraphics[width=1\linewidth]{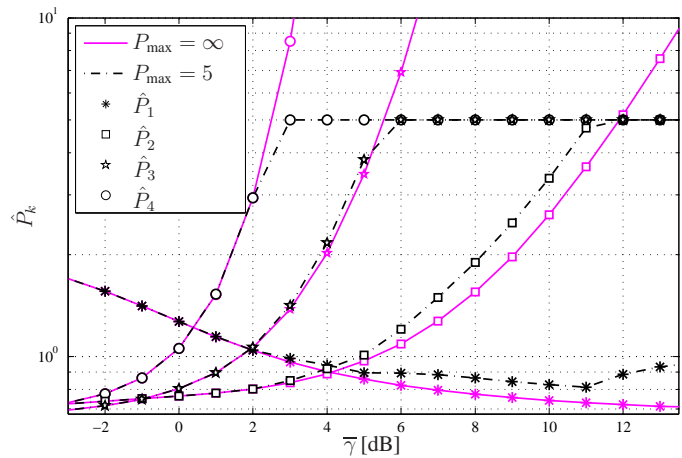}}
\\
(b)
\caption{Optimized allocation policies $\hat{P}_{k}$ as a function of $\ov{\gamma}$  in the case of (a) IR-HARQ and (b) CC-HARQ. $ K=4$, $m=2$, $R=1.5$. Both cases of unbounded (i.e., $P_{\tr{max}}=\infty$) and bounded peak power (i.e., $P_{\tr{max}}=5$) are shown for comparison.} \label{power.IR.2}
\end{center}
\end{figure}

 
For the case of $m=2$, Fig.~\ref{fig.outage.IR}a and Fig.~\ref{fig.outage.IR}b  present the optimized outage probability in the case of \gls{irharq} and \gls{ccharq}, respectively. We show both cases when $ P_{\tr{max}}=5$ and $ P_{\tr{max}}= \infty$. We also plot the outage probability of constant-power transmissions (CO). We can see, that for high SNR, the optimized results outperform CO HARQ, which is due to the increases diversity of both power adaptation and allocation strategies. On the other hand, the results for CO HARQ can outperform AL HARQ because the latter is based on the approximations, which loose their validity for low SNR. For example, for $\ov{\gamma}<-2$  dB in Fig.~\ref{fig.outage.IR}a and for $\ov{\gamma}<0$  dB in Fig.~\ref{fig.outage.IR}b.
When $P_{\tr{max}}=5$ instead of $ P_{\tr{max}}=\infty$, the gain of the optimized outage compared with CO starts to decrease after a specific value $\ov{\gamma}_\tr{0}$ of the average SNR $\ov{\gamma}$. For example, in the case of allocation  with $K=4$, $\ov{\gamma}_{0}\approx0$ dB for \gls{irharq} and $\ov{\gamma}_\tr{0}\approx2$ dB for \gls{ccharq}. This is justified by the fact that the constraint $P_{\tr{max}}= 5$ becomes active for $\ov{\gamma} \geq \ov{\gamma}_{0}$, as can be seen in Fig.~\ref{power.IR.2}. Moreover, when the maximum power constraints are active, the diversity of the adaptation/allocation schemes is the same as the diversity of the constant power transmission.
 
\begin{figure}[h!]
\psfrag{xlabel}[c][c][\siz]{$\ov{\gamma}$\! {[dB]}}
\psfrag{ylabel}[c][c][\siz]{$f_{K}$}
\psfrag{m=1}[l][l][\siz]{$m=1$}
\psfrag{m=2}[l][l][\siz]{$m=2$}
\psfrag{m=3}[l][l][\siz]{$m=3$}
\psfrag{K=2}[l][l][\siz]{$K=2$}
\psfrag{K=4}[l][l][\siz]{$K=4$}
\psfrag{DP Alloc unbbbbbb}[l][l][\siz]{AL with $P_{\tr{max}}=\infty$ }
\psfrag{DP Adap unb}[l][l][\siz]{AD with $P_{\tr{max}}=\infty$ }
\begin{center}
\scalebox{0.5}{\includegraphics[width=1\linewidth]{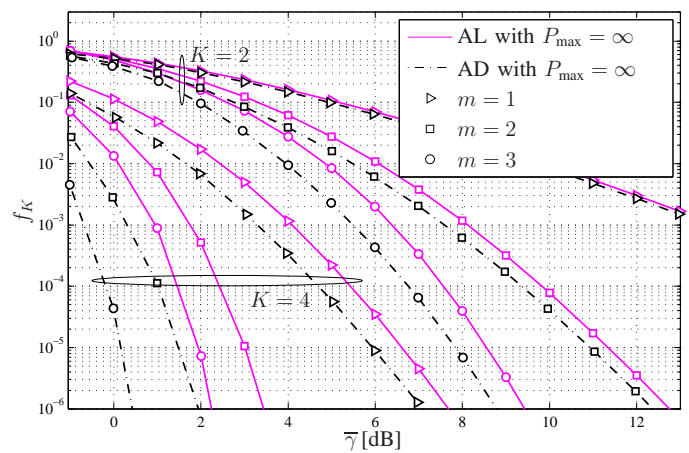}}
\caption{Optimized outage probability when using the optimized power adaptation (AD) and allocation (AL) policies in the case of CC-HARQ when $K=2,4$ and all variables correspond to Nakagami-$m$ fading with $m=1, 2, 3$, $R=1.5$, and $P_{\tr{max}}=\infty$.} \label{CC_outagP_2}
\end{center}
\end{figure}

\begin{figure}[h!]
\psfrag{xlabel}[c][c][\siz]{$\ov{\gamma}$\! {[dB]}}
\psfrag{ylabel}[c][c][\siz]{$f_{K}$}
\psfrag{m=1}[l][l][\siz]{$m=1$}
\psfrag{m=2}[l][l][\siz]{$m=2$}
\psfrag{m=3}[l][l][\siz]{$m=3$}
\psfrag{DP}[l][l][\siz]{DP}
\psfrag{GP}[l][l][\siz]{GP}
\begin{center}
\scalebox{0.5}{\includegraphics[width=1\linewidth]{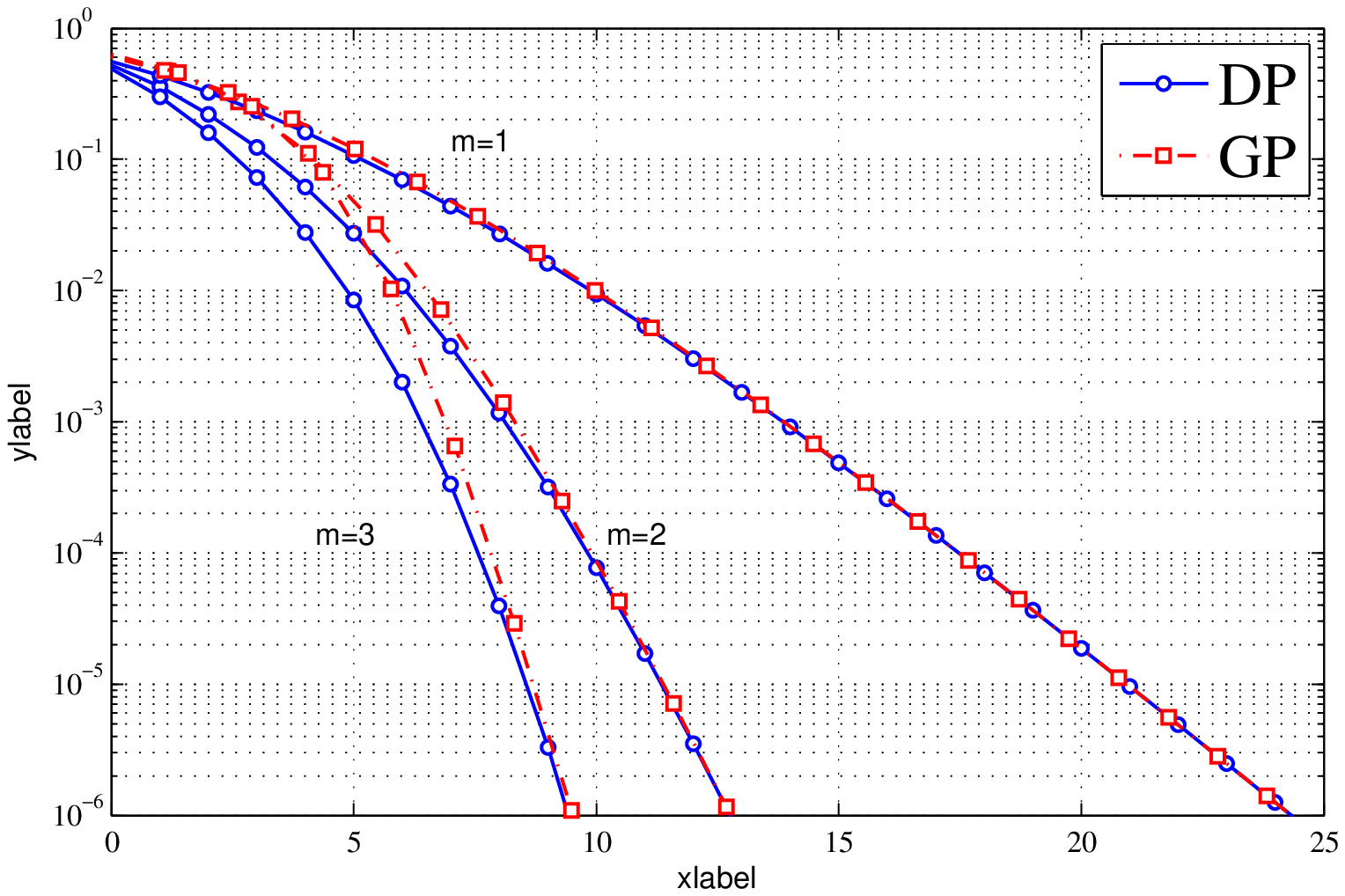}}
\\
(a)
\\
\scalebox{0.5}{\includegraphics[width=1\linewidth]{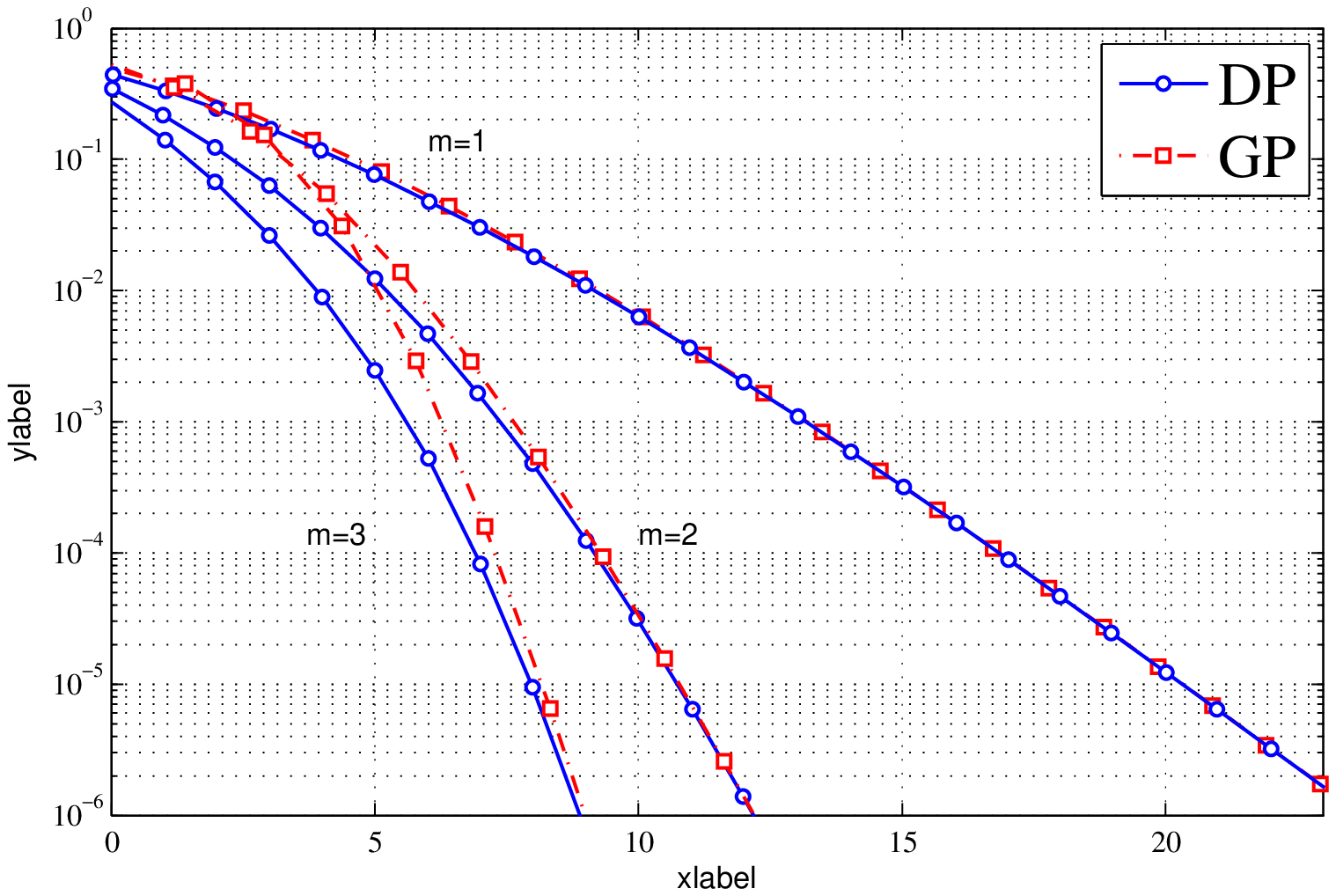}}
\\
(b)
\caption{Outage probability when HARQ process use the optimized allocation power policies find by dynamic programming (DP) and geometric programming (GP)  in the case of (a) CC-HARQ and (b) IR-HARQ. $K=2$ and all variables correspond to Nakagami-$m$ fading with $m=1, 2, 3$ and $R=1.5$.}\label{fig:GP-MRC}
\end{center}
\end{figure}

Fig.~\ref{CC_outagP_2} illustrates the gain of the power adaptation over allocation strategies, where it is clear that the gain is not only a function of the maximum number of transmissions $K$, but is also depends on the channel parameter $m$. In particular, for $K=2$ we obtain the gain of approximately  0.1 dB, 0.2 dB, and  0.5 dB for $m=1$, $m=2$, and $m=3$, respectively. For $K=4$, the respective gains increase and also grow with $m$; they are approximately given by  0.5 dB, 1.5 dB, and 1.8 dB.
In Fig.~\ref{fig:GP-MRC}, we compare the optimized solutions obtained using \gls{dp} and \gls{gp} for \gls{ccharq} and \gls{irharq}. For high SNR, and as expected, the solutions of GP converge to the optimized solution obtained with DP.

%
\section{Conclusion}\label{Sec:Conclusions}
In this paper, we analyzed the impact of multi-bit feedback on the performance of HARQ protocols in terms of outage probability. We analyzed HARQ with Chase combining and Incremental Redundancy in Nakagami-$m$ block fading channels. We show that an optimized power allocation/adaptation strategy throughout the transmissions leads to notable gains over the power-constant HARQ. Adding multi-bit feedback improves the performance and the achievable gains increase with the allowed number of transmissions as well as with the parameter $m$ of the Nakagami-$m$ distribution.


%
\section*{Appendix A}
\label{Sec:AppendixA}
We aim to determine the expression of $h_{k}$ and $A_{K}$ (required in \eqref{A.S.R} and \eqref{eq:A.f.K.Al} respectively) for the case of \gls{irharq} and \gls{ccharq}. For that, we will derive a simple and accurate approximation of $f_{k}$ defined in \eqref{eq.outage}. Clearly, calculating the outage probability in \eqref{eq.outage} for the power allocation scheme requires the derivation of the \gls{cdf} of the sum of $k$ independent random variables: \mbox{$C_{k}=\log_{2}\left(1+\gamma_{k}\hat{P}_{k}\right)$} in the case of \gls{irharq}  and $\sigma_{k}=\gamma_{k}\hat{P}_{k}$ in the case of \gls{ccharq}.
\subsection{CC-HARQ}

We use a simple and accurate method to evaluate the outage probability at the output of \gls{mrc} receivers in arbitrarily fading channels introduced in \cite{jabi12}. The approximation is based on the so-called \gls{spa} \cite{Lugannani80,Butler07_book}. For the special case of Nakagami-$m$ fading channels, the outage probability can be approximated by \cite{jabi12}
\begin{align} \label{eq:appr.outage.CC} 
f_{K}&\approx  \left (\frac{\exp(1) \cd \gamma_{\tr{th} }}{{\tilde{m}_{K}}}\right)^{\!\tilde{m}_{K}} \cd \frac{1}{\sqrt{2\pi {\tilde{m}_{K}}}} \cd \prod_{k=1}^{K} \left( \frac{m_{k}}{\ov{\gamma}_{k}  \hat{P}_{k}} \right)^{\!m_{k}},\\
&=A_{K}^{\gls{cc}}\cd \prod_{k=1}^{K}  \frac{1}{\hat{P}_{k}^{m_{k}}},
\end{align}
where  $\tilde{m}_{K}=\sum_{k=1}^{K}m_{k}$ and
\begin{align} \label{eq_A_CC}
A_{K}^{\gls{cc}}=\left (\frac{\exp(1) \cd \gamma_{\tr{th} }}{{\tilde{m}_{K}}}\right)^{\!\tilde{m}_{K}} \cd \frac{1}{\sqrt{2\pi {\tilde{m}_{K}}}} \cd \prod_{k=1}^{K} \left( \frac{m_{k}}{\ov{\gamma}_{k} } \right)^{\!m_{k}}.
\end{align}

In this case, we can easily show that $h_{k}$, required in \eqref{A.S.R}, is given by 
\begin{align} \label{eq.outage.appendix} 
h_{k}
&=\begin{cases}
\left (\frac{\exp(1) \gamma_{\tr{th}} m_{k}}{\ov{\gamma}_{k}\tilde{m}_{k}}  \right)^{\!m_k} \cd \left( 1-\frac{m_{k}}{\tilde{m}_{k}}\right)^{\!\tilde{m}_{k-1}+0.5},&\text {for}~2 \leq k \leq K \\\\
\frac{1}{\sqrt{2\pi m_{1}}}\left(\frac{\exp(1) \cd \gamma_{\tr{th}} }{\ov{\gamma}_{1}}  \right)^{\!m_1},&\text{for}~k=1 \\
\end{cases}.
\end{align}

\subsection{IR-HARQ}

In \cite{laneman03_01} and \cite{Lee10}, the authors develop a simple and powerful way of characterizing the performance of diversity schemes via limiting analysis of outage probabilities in Rayleigh fading channels. We use a similar analysis to approximate the outage probability in the case of \gls{irharq} for Nakagami-$m$ fading channels. The key idea is the following:
\begin{Theorem} \cite[Theorem.~1]{laneman03_01} \cite[Lemma.~1]{Lee10}\label{theo1}\\
Let $Z$ and $W$ be two independent random variables. If their CDF verify
\begin{align}
\lim_{\ov{\gamma}\rightarrow \infty} \ov{\gamma}^{n_{1}}\cd \cdf_{Z}{(t)}=a\cd q(t), 
\end{align}
\begin{align}
\lim_{\ov{\gamma}\rightarrow \infty} \ov{\gamma}^{n_{2}}\cd \cdf_{W}{(t)}=b\cd g(t),
\end{align}
where $n_{1}$, $n_{2}$, $a$ and $b$ are constants, $g(t)$ and $q(t)$ are monotonically increasing functions, and the derivative of $q(t)$ (denoted as $q'(t)$) is integrable, then the CDF of the sum $Y=Z+W$ satisfies
\begin{align}
\lim_{\ov{\gamma}\rightarrow \infty} \ov{\gamma}^{n_{1}+n_{2}} \cdf_{Y}{(t)}=ab \cd \int_{0}^{t} g(x)\cd q'(t-x) \tr{d}x.
\end{align}
\end{Theorem} 

Since $\gamma_{k}$ follows a gamma distribution, it is easy to show that
\begin{align}
\lim_{\ov{\gamma}\rightarrow \infty} \ov{\gamma}^{m_{k}} \cdf_{C_{k}}(t)=\frac{m_{k}^{m_{k}}}{\hat{P}_{k}^{m_{k}}\Gamma(m_{k}+1)}\cd \left(2^{t}-1\right)^{m_{k}}.
\end{align}

\begin{figure}[th]
\psfrag{xlabel}[c][c][\siz]{$\ov{\gamma}$\! {(dB)}}
\psfrag{ylabel}[c][c][\siz]{$f_{K}$}
\psfrag{K=1}[l][l][\siz]{$K=1$}
\psfrag{K=2}[l][l][\siz]{$K=2$}
\psfrag{K=3}[l][l][\siz]{$K=3$}
\psfrag{K=4}[l][l][\siz]{$K=4$}
\psfrag{EX}[l][l][\siz]{$f_{K}$}
\psfrag{AP}[l][l][\siz]{$\hat{f}_{K}$}
\begin{center}
\scalebox{0.5}{\includegraphics[width=1\linewidth]{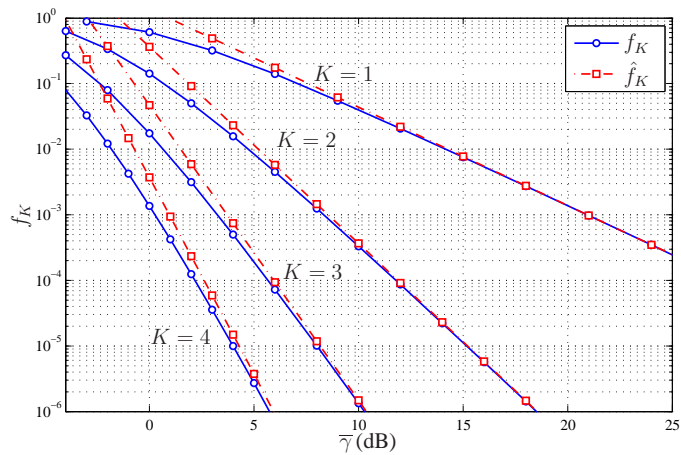}}
\caption{Exact outage probability $f_{K}$ compared with the SPA approximation $\hat{f}_{K}$ \eqref{eq:appr.outage.IR} in the case of IR-HARQ for Nakagami-$m$ when $ K=1,2, 3, 4$,  $R=1$ and $m=1.5$.}\label{fig:app:out}
\end{center}
\end{figure}

Using \theref{theo1} with $t =R$ recursively, we have an approximation of the outage probability as
\begin{align} \label{eq:appr.outage.IR}
f_{K}\approx \hat{f}_{K}&= g_{K}(R) \cd \prod_{k=1}^{K} \frac{m_{k}^{m_{k}}}{\hat{P}_{k}^{m_{k}}\ov{\gamma}^{m_{k}}\Gamma(m_{k}+1)},\\
&=A_{K}^{\gls{ir}} \cd \prod_{k=1}^{K}  \frac{1}{\hat{P}_{k}^{m_{k}}},
\end{align} 
where
\begin{align}
g_{k}(t)=\int_{0}^{t} g_{k-1}(x) q'_{k}(t-x) \tr{d}x,
\end{align} 
with $g_{0}=1$ and $q_{k}(t)=\left(2^{t}-1\right)^{m_{k}}$. Thus, $A_{K}^{\gls{ir}}$ is given by
\begin{align} \label{eq_A_IR}
A_{K}^{\gls{ir}}=g_{K}(R) \cd \prod_{k=1}^{K} \frac{m_{k}^{m_{k}}}{\ov{\gamma}^{m_{k}}\Gamma(m_{k}+1)},
\end{align}
where $g_k( R)$ can be calculated numericaly. We show the accuracy of the approximation in Fig. \ref{fig:app:out}.

Since $g_k(t)$ are independent of the transmitted power and/or SNR, the expression of $h_{k}$, required in \eqref{A.S.R},  is thus given by 
\begin{align}
h_{k}&=\begin{cases}
\displaystyle{\frac{g_{k}(R)}{g_{k-1}(R)}\cd \frac{m_{k}^{m_{k}}}{\ov{\gamma}^{m_{k}}\Gamma(m_{k}+1)}},& 2 \leq k \leq K\\
\displaystyle{ \left({\frac{m_{1}(2^R-1)}{\ov{\gamma}}}\right)^{m_{1}}\cd \frac{1}{\Gamma(m_{1}+1)} },&k=1\\
\end{cases}.
\end{align}

\section*{Appendix B}\label{Sec:AppendixB}

In this appendix, we show how we solve the optimization problem \eqref{eq:GP.op} written as a \gls{gp} problem \cite{GPbook} , \cite{Luptacik} , \cite{Boyd04_Book} in the standard primal form
\begin{align} \label{eq:GP.op1}
\underset{\mb{x}}{\min} \Big\{ g_{0}(\mb{x})= A_{K} \cd \prod_{k=1}^{K} x_{k}^{-m} \Big\} ~~~\tnr{s.t.} ~g_{1}(\mb{x})=\sum_{k=1}^{K} A_{k-1}  \cd {x}_{k}  \prod_{l=1}^{k-1} {x}_{l}^{-m}  \leq 1,
\end{align}
where $\mb{x} =[x_{1}, x_{2},\ld,x_{K}]$.
The dual problem corresponding to the primal problem \eqref{eq:GP.op1} is defined as 
{\small
\begin{align}
&~~~~\underset{\mb{\delta}}\max ~ v (\mb{\delta})=\left(\lambda(\mb{\delta}) \right)^{\lambda(\mb{\delta})} \left( \frac{A_{K}}{\delta_{1}} \right)^{\delta_{1}} \prod_{i=2}^{K+1} \left( \frac{A_{i-2}}{\delta_{i}} \right)^{\delta_{i}} ,\label{eq:GP.dual}\\
&\tnr{s.t.}\quad \begin{cases} \label{eq:dual.constr} 
\delta_{1}=1,\\
\delta_{i} \geq 0, &\forall i\in\{1,..,K+1\} \\ 
-m\cd \delta_{1}+ \delta_{j} -m\cd \sum_{i=j+1}^{K+1}\delta_{i}=0,&\forall j\in\{2,..,K+1\} 
\end{cases},
\end{align}
}
where $\boldsymbol{\delta}=[\delta_1,\ld, \delta_{K+1}]$ and 
\begin{align}\label{eq:lambda}
\lambda(\mb{\delta})=\sum_{i=2}^{K+1} \delta_{i}.
\end{align}
This is a GP with a zero degree of difficulty \cite{Luptacik}, which implies that the unique solution $\mb{\delta}^{*}$ of the dual constraints \eqref{eq:dual.constr} is also the solution of \eqref{eq:GP.dual} .  Because the dual constraints are linear,  $\mb{\delta}^{*}$ can be determined easily by solving \eqref{eq:dual.constr} as
\begin{align} 
\delta_{i}^{*}=
\begin{cases} 
1, & \text{if} ~ i=1\\
m\cd (m+1)^{(K+1-i)}, & \text{if} ~ i\in\{2,...,K+1\}
\end{cases}.
\end{align}
Defining $\mb{x}^{*}$ as the argument which maximizes \eqref{eq:GP.op1}, the optimal solution of \eqref{eq:GP.op1} is given by \cite[pp.~114-116]{GPbook}
\begin{align}\label{eq:GP.Sol}
g_{1}(\mb{x}^{*})=v (\mb{\delta}^{*})=\left(\lambda(\mb{\delta}^{*}) \right)^{\lambda(\mb{\delta}^{*})} \cd A_{K} \cd \prod_{i=2}^{K+1} \left( \frac{A_{i-2}}{\delta_{i}^{*}} \right)^{\delta_{i}^{*}},
\end{align}
if and only if
\begin{align} \label{eq:x1.GP}
x_{i}^{*}=\begin{cases}
\displaystyle{\frac{{\delta}_{2}^{*}}{\lambda(\mb{\delta}^{*}) \cd A_{0}}},& \text{if}~ i=1\\
\displaystyle{\frac{{\delta}_{i+1}^{*}}{\lambda(\mb{\delta}^{*}) \cd A_{i-1}\cd \prod_{j=1}^{i-1} \left(x_{j}^{*}\right)^{-m}}}, & \text{if}~i\in\{2,..,K\}
\end{cases},
\end{align}
where
\begin{align}
\lambda(\mb{\delta}^{*})=(m+1)^{K}-1.
\end{align}
Basically, \eqref{eq:x1.GP} is the closed-form expression of the optimal power policy.

\balance



\bibliographystyle{IEEEtran}

\end{document}